\documentstyle[graphicx,pre,multicol,aps]{revtex}
\begin{document}
\title{Self-written waveguides in photopolymerizable resins}

\author{Satoru Shoji and Satoshi Kawata}
\address{Department of Applied Physics, Osaka University, 
Yamadaoka 2-1, Suita, Osaka 565-0871, Japan}

\author{Andrey A. Sukhorukov and Yuri S. Kivshar}
\address{Nonlinear Physics Group, 
Research School of Physical Sciences and Engineering,\\
Institute for Advanced Studies,
Australian National University, Canberra, ACT 0200, Australia}

\maketitle
\begin{abstract}
We study the optically-induced growth and interaction of self-written waveguides in a photopolymerizable resin. We investigate experimentally how the interaction depends on the mutual coherence and relative power of the input beams, and suggest an improved analytical model that describes the growth of single self-written waveguides and the main features of their interaction in photosensitive materials.
\end{abstract}


\vspace*{-1cm}
\begin{multicols}{2}
\narrowtext

Optical self-action effects occur when the beam induces a refractive index change in the medium through which it propagates. Such effects are usually associated with the generation, propagation, and interaction of {\em spatial optical solitons}~--- the self-trapped optical beams that exist due to the balance between diffraction and nonlinearity~\cite{soliton}. 

One of the important concepts that drive the research on optical solitons is their possible use as steerable self-induced waveguides that can guide other beams of different polarization or wavelength. However, in order to "freeze" the soliton-induced waveguides, one should use {\em photosensitive materials} which experience long-lasting refractive index changes in response to illumination at specific wavelength~\cite{monro}. The main question then is if the properties of such self-written waveguides are similar to the properties of spatial solitons. This is especially important for writing multiple waveguides and optical components, such as Y- and X-junctions, based on the intersection of two (or more) self-written waveguides.

In this Letter we study, first experimentally and then theoretically, the optically-induced growth and interaction of self-written waveguides in photopolymerizable resins, and demonstrate both differences and similarities between the interaction of spatial solitons and self-written waveguides.

As a photosensitive optical material, we use a photopolymerizable resin (PR) where single and multiple self-written waveguides (``fibers'') can be grown by a one-photon absorption process~\cite{experiment}. First, we reproduce the experimental results on the growth of single waveguides earlier reported in Ref.~\cite{experiment}, and then we study experimentally different regimes of interaction between two self-written waveguides. For the experiments, we use PR in the form of a liquid urethane acrylate photopolymer (SCR-500: Japan Synthetic Rubber Co., Ltd). The PR is filled in a glass cell and the beam is focused on the entrance face of the cell. The PR initial refractive index (before the illumination) is 1.53, which increases to 1.55 gradually with the photo-polymerization reaction. In order to study the beam collision, a light beam operated from a He-Cd laser ($\lambda=441.6\; {\rm nm}$) is split into two beams, which are then focused onto the resin, simultaneously or with a delay. 

The optical axes of two beams intersect inside the PR so that the two waveguides growing simultaneously from the two beam spots collide with each other inside the photosensitive material. The polymerized structure is observed by a charge-coupled-device (CCD) camera from a side of the sample cell.

\begin{figure}[H]
\centerline{\includegraphics[width=8cm,clip]{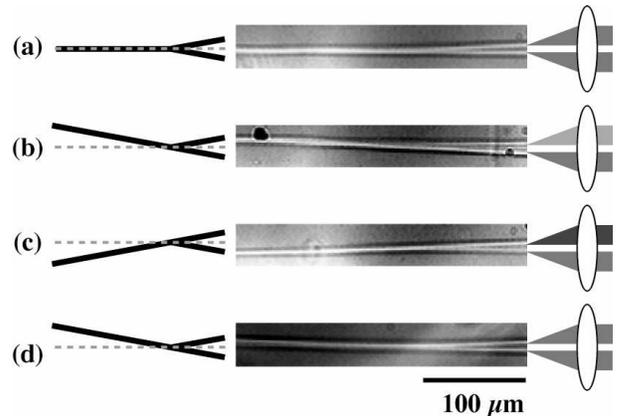}}
\caption{ \label{fig:experiment}
Experimental data showing the collisions between two self-written waveguides. The power of the lower beam is $0.1\,{\rm mW}$; the power of the upper beam is (a,d)~$0.1\,{\rm mW}$, (b)~$0.07\,{\rm mW}$, and (c)~$0.13\,{\rm mW}$. The time delay between the lower and upper beams is (a-c)~$0\;{\rm sec}$, and (d)~$2\;{\rm sec}$.
}
\end{figure}

Figure~\ref{fig:experiment}(a) shows that when two self-written waveguides collide with each other simultaneously, they can merge to form a single waveguide.  The power of each input beam is $0.1\,{\rm mW}$, the exposure time is $2\,{\rm sec}$, and the input beam width is $\simeq 0.96\,{\rm \mu m}$ (the numerical aperture of the focusing lens is $0.23$). Similar to the interaction of two solitons~\cite{soliton}, the merging of self-growing waveguides strongly depends on the collision angle between two waveguides, as was discovered in Ref.~\cite{experiment}. In the PR we used, the merging does not occur when the collision angle is larger than $9^\circ$.  For all the experiments reported in this Letter, the collision angle is $6.4^\circ$, which is lower than the critical value, so that the growing waveguides always merge after collision. In the case of Fig.~\ref{fig:experiment}(a), the waveguide after merging is growing along the direction that bisects the angle between two waveguides which formed before the collision, because the input powers of two beams are symmetric and balance each other.

Figures~\ref{fig:experiment}(b,c) show that the variation of the power ratio of two input beams can change the growing direction of the waveguide after merging. In Figs.~\ref{fig:experiment}(b,c), one input beam power is fixed at $0.1\;{\rm mW}$, and the other input beam power is changed to $0.07\;{\rm mW}$ (b) and $0.13\;{\rm mW}$ (c). The collision angle is $6.4^\circ$, the exposure time and input beam width are the same as in the experiment shown in Fig.~\ref{fig:experiment}(a). The growing direction of the merged waveguide is tilted towards the optical axis of the beam with higher input power, in comparison to the case of the symmetric input powers.
The similar effect is observed when two beams are launched one after another, as shown in Fig.~\ref{fig:experiment}(d). In this case, the resulting self-written waveguide follows the direction of the beam that enters the medium first. This effect is in a sharp contrast with the interaction of spatial solitons of unequal amplitudes~\cite{soliton}.

In order to explain these experimental results, as well as the specific features observed in the growth of single self-written waveguides earlier reported in Ref.~\cite{experiment}, we should employ an appropriate model that describes the growth of self-written waveguides in a photosensitive optical material. Since the polymerization is a slow process compared to the optical beam propagation through a photosensitive medium, at any given time the light distribution can be described by a ``stationary'' wave equation for the electric field envelope~\cite{monro,yariv}. Since the variation of the refractive index is small, the wave equation can be simplified by applying the paraxial approximation,
\begin{equation} \label{eq:nls}
   i \frac{\partial E({\bf r}, z, \tau)}{\partial z} 
   + D \nabla_{\bot}^2 E
   + \nu({\bf r}, z, \tau) E
   = 0 ,
\end{equation}
where $E  = {\cal E}/ {\cal E}_0$ is the normalized complex electric field envelope, $\nabla_{\bot}^2$ is the Laplace operator in the transverse coordinate system (${\bf r}$), and the dimensionless variables are the following: $z = Z / a$ is the propagation distance, ${\bf r} = {\bf R} / a$ is the spatial coordinate in the transverse  cross-section, $\tau = t / t_0$ is time, $D = (2 a k_0 n_0)^{-1}$ is the diffraction coefficient, and $\nu = \Delta n\,a k_0$. Here $a$ is the characteristic beam width, $t_0$ defines the time scale, $k_0 = 2 \pi / \lambda$ is the wave number in vacuum, $n_0$ is the initial refractive index of the resin, and $\Delta n(z, {\bf r}, \tau)$ is the refractive index change due to polymerization. We define the normalized power as $P = \int |E|^2 d{\bf r}$, which is a conserved quantity of Eq.~(\ref{eq:nls}).

The refractive index and the corresponding light distribution change slowly as the medium density increases due to photoinduced polymerization. The polymerization reaction is an irreversible multi-step process, involving the creation of radicals from initiators by means of one-photon absorption, their interaction with monomers, and combination of monomers into polymer chains. When the medium is completely polymerized, the refractive index reaches its maximum value. Since the number of created radicals is proportional to the number of absorbed photons, the rate of the polymerization process is determined by the light intensity. 
The polymerization reaction is initially delayed (see, e.g., Ref.~\cite{yariv}), while radicals are scavenged by the highly reactive oxygen molecules present in the resin. Moreover, in the case of low rates of one-photon absorption, the delay time increases for smaller light intensities~\cite{3d1p}. Therefore, for a given exposure time, the polymerization can only occur in the regions where the light intensity exceed a certain {\em threshold}. Although the accurate description of the polymerization process in resins should involve a solution of the kinetic equations (see, e.g., Ref.~\cite{rate} for the case when the oxygen concentration is negligible), earlier studies indicate that acceptable results can be obtained by employing simpler phenomenological models~\cite{monro}. Therefore, we modify the model introduced earlier~\cite{yariv} to account for the threshold-type polymerization rate dependence on the light intensity, and define the temporal evolution of the refractive index change $\Delta n({\bf r}, z, \tau)$ as follows
\begin{equation} \label{eq:dn}
   \frac{\partial \Delta n}{\partial \tau} = 
             A \left( 1 - \frac{\Delta n}{\Delta n_s} \right)
             \left\{ \begin{array}{l}
                    |E|^2 - I_{\rm th},\, |E|^2 \ge I_{\rm th}, \\
                    0,\, |E|^2 < I_{\rm th},
               \end{array} \right.
\end{equation}
where, for simplicity, we neglect a temporal delay for the intensities above the threshold. Here $\Delta n_s$ is the maximum refractive index change, $I_{\rm th}$ is the normalized threshold intensity, and $A$ is a coefficient characterizing the polymerization rate.

\begin{figure}[H]
\vspace*{-5mm}
\centerline{\includegraphics[width=8cm,clip]{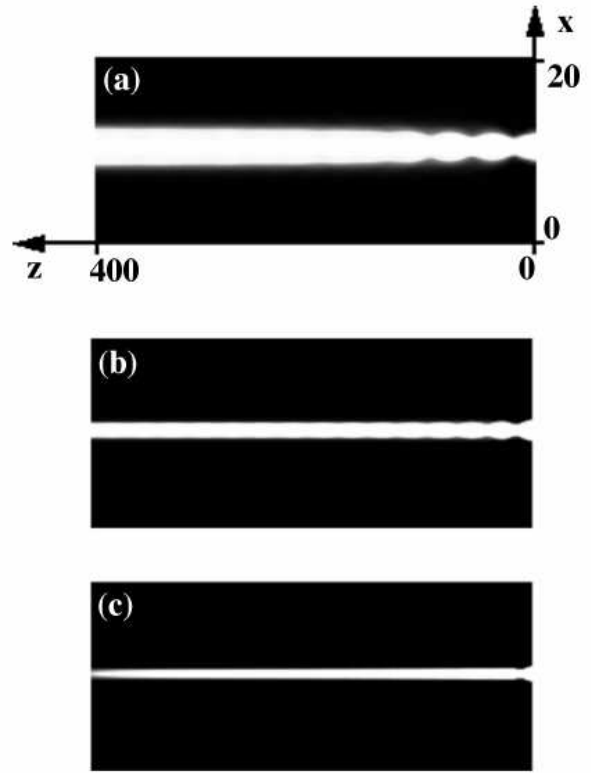}}
\caption{ \label{fig:theory-growth}
Numerically calculated changes in the refractive index profile for different values of the intensity threshold $I_{th}$: (a)~$0$, (b)~$0.01$, and (c)~$0.025$. For all the plots, the input power is $P \simeq 0.1$, exposure time $\tau = 1200$, beam width ${\rm FWHM} \simeq 1.5$, and the model parameters are $A = 0.0025$, $n_0 = 1.53$, $\Delta n_s = 0.02$, $t_0 = 1 {\rm m s}$, and $a = 1\;{\rm \mu m}$.
}
\end{figure}

In order to describe the qualitative features of the self-written waveguides in a photopolymerizable resin, we make a further simplification by considering a (1+1)-dimensional spatial geometry. Analysis of the full (2+1)-dimensional equations will be presented elsewhere. We perform numerical simulations for the input Gaussian beams, $E_0(x) = E_m \exp(- x^2 / x_0^2)$, so that the full-width at the half-maximum (FWHM) of intensity distribution is $x_0 \sqrt{2 \ln 2}$, and the normalized power is $P = E_m^2 x_0 \sqrt{\pi/2}$. Refractive index profiles corresponding to the self-growing waveguides are shown in Figs.~\ref{fig:theory-growth}(a-c). For comparison with the earlier studies~\cite{monro,yariv}, we first consider the case of zero threshold ($I_{\rm th} = 0$). As shown in Fig.~\ref{fig:theory-growth}(a), the corresponding waveguide becomes broader away from the input face, developing ``eyes'' at the input, as is observed in photosensitive glass~\cite{monro}. However these features are not observed in a resin. In contrast, the waveguide width is stabilized when the threshold is taken into account, see Figs.~\ref{fig:theory-growth}(b,c). This is exactly the behavior which is observed experimentally in photopolymerizable resins~\cite{experiment}.

Based on the numerical results, we seek stationary solutions of Eqs.~(\ref{eq:nls}) and~(\ref{eq:dn}) describing self-written waveguides with constant width away from the input face. Due to the saturation effect, the profile of a single waveguide is, up to a constant shift in the transverse direction,
\begin{equation} \label{eq:prof}
    \Delta n(x) = \left\{  \begin{array}{l} 
                         \Delta n_s, \, |x| < d/2 ,\\
                         0, \, |x| \ge d/2 ,
                 \end{array} \right. 
\end{equation}
where $d$ is the width. On the other hand, the optical modes of this waveguide should satisfy the self-consistency relations following from Eq.~(\ref{eq:dn}), i.e. $|E(x)|^2 < I_{\rm th}$ for $|x| > d/2$ and $|E(x)|^2 > I_{\rm th}$ for $|x| < d/2$. These relations uniquely define the optical field profile if asymmetric modes are not excited, and the waveguide supports only a single symmetric mode, which is the case for $d < 2 \pi \sqrt{D/\nu_s}$,
\[
    E(x,z) = \sqrt{I_{\rm th}} e^{i \beta z} 
              \left\{  \begin{array}{l} 
                      \cos( q x ) / \cos( q d/2 ) , \, |x| < d/2 ,\\
                      \exp[ - p (|x| -d/2)], \, |x| \ge d/2 ,
                 \end{array} \right. 
\]
where $\nu_s = \nu(\Delta n_s)$, $q = \sqrt{(\nu_s-\beta)/D}$, and $p = \sqrt{\beta/D}$. The propagation constant $\beta$ should be chosen to satisfy the continuity of the solution $E(x,z)$, which leads to the condition $\tan(q d/2) = p/q$. Then, we find that a waveguide supporting a single symmetric mode can be formed if $I_{\rm th} < I_{\rm max} < 7\, I_{\rm th}$. These analytical results explain the stabilization of the waveguide width and suppression of beating as the threshold is increased in numerical simulations. We have also checked that only the fundamental symmetric mode is excited in the waveguide shown in Fig.~\ref{fig:theory-growth}(c), away from the input face.

\begin{figure}[H]
\vspace*{-5mm}
\centerline{\includegraphics[width=8cm,clip]{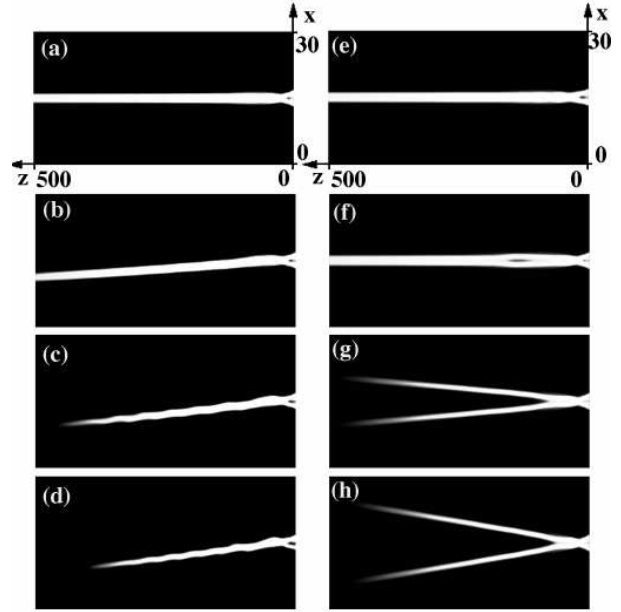}}
\caption{ \label{fig:theory-collis}
Numerically calculated changes in the refractive index profile for colliding waveguides, formed by inclined input beams. 
In Figs.~(a-c) the collision angle is $1^\circ$, the power of the upper input beam is $P=0.1$, the lower beam power is (a)~$P = 0.1$, (b)~$0.092$, (c)~$0.077$, and (d)~$0.063$.
In Figs.~(e-h) the power of both beams is $P=0.1$, and the collision angle is (e)~$1^\circ$, (f)~$2^\circ$, (g)~$2.5^\circ$,  (h)~$3^\circ$.
For all the plots $I_{\rm th} = 0.025$, and other parameters are the same as in Fig.~\ref{fig:theory-growth}.
}
\end{figure}

We now consider interaction between two waveguides formed by inclined input beams. First, we study the effect of the varying power ratio between the two beams. As is shown in Figs.~\ref{fig:theory-collis}(a-d), after the collision point the waveguide direction is determined by the more powerful beam. This feature agrees well with the experimental observations presented above. Next, we study the dependence on the collision angle and find that the waveguides can merge only at small collision angles, but pass through each other when the collision angle is increased above some threshold, see Figs.~\ref{fig:theory-collis}(e-h). This result is also similar to the reported experimental data. 

In conclusion, we have suggested an improved analytical model that describes the growth and interaction of self-written waveguides in photopolymerizable resins, and demonstrated both differences and similarities between incoherent interaction of spatial optical solitons and intersection of self-written waveguides. Our results may be useful for a design of optical components and structures based on the waveguides self-written in photosensitive materials.

The work has been partially supported by the Australian Photonics Cooperative Research Centre, the Japanese Society for Promotion of Science, and the Australian Academy of Science.

\end{multicols}

\vspace*{-5mm}
\begin{thebibliography}{99}
\vspace*{-15mm}

\bibitem{soliton}
See, e.g., 
G.~I.~Stegeman and M.~Segev,
Science {\bf 286}, 1518 (1999),
and references therein.

\bibitem{monro}
See the recent comprehensive review by 
T. M. Monro, C. M. De Sterke, and L. Poladian,
J. Mod. Opt. {\bf 48}, 191 (2001),
and references therein.

\bibitem{experiment}
S. Shoji and S. Kawata, 
Appl. Phys. Lett. {\bf 75}, 737 (1999).

\bibitem{yariv}
A.~S. Kewitsch and A. Yariv, 
Opt. Lett. {\bf 21}, 24 (1996);
Appl. Phys. Lett. {\bf 68}, 455 (1996).

\bibitem{3d1p}
S. Maruo and K. Ikuta, 
Appl. Phys. Lett. {\bf 76}, 2656 (2000).  

\bibitem{rate}
See, e.g.,
D. Engin, A.~S. Kewitsch, and A. Yariv,
J. Opt. Soc. Am. B {\bf 16}, 1213 (1999),
and references therein.

\end{thebibliography}
\end{document}